\documentclass[authoryear,preprint,review,12pt]{elsarticle}

\usepackage{graphics}

\usepackage{graphicx}

\usepackage{epsfig}

\usepackage{amssymb}

\usepackage{amsthm}

\usepackage{lineno}

\journal{New Astronomy}

\begin{document}

\begin{frontmatter}



\title{An Eccentric Eclipsing Binary: CG\,Aur}


\author{E. Sipahi\corref{cor1}}
\ead{esin.sipahi@mail.ege.edu.tr}
\cortext[cor1]{Corresponding author}

\author{H. A. Dal}

\address{Ege University, Science Faculty, Department of Astronomy and Space Sciences, 35100 Bornova, \.{I}zmir, Turkey}

\begin{abstract}

In this study, we present CG\,Aur's photometric observations obtained in the observing seasons 2011 and 2012, the first available multi-colour light curves. Their shape indicates that the system is an Algol binary. The light curve analyses reveal that CG\,Aur is a detached binary system with an effective temperature difference between the components, approximately 1000 K. The first estimate of the absolute dimensions of the components indicated that the system locates on the main sequence in the HR diagram. The primary component is slightly evolved from the ZAMS.
\end{abstract}


\end{frontmatter}

% \linenumbers


\section{Introduction}
CG\,Aur ($V=11^{m}.43$) is classified as an eclipsing binary in the SIMBAD database. The system was first discovered by \citet{Hof35}, while \citet{Kan39} gave the first light elements. The photographic light curve of the system was obtained by \citet{Kur51}. The system was observed by \citet{Zak95} and the colour indexes of $U-B$ and $B-V$ were given as 0$^{m}$.36 and 0$^{m}$.66 in this study, respectively. Presenting the $O-C$ analyses of the system, \citet{Wol11} indicated that CG\,Aur is an interesting triple and eccentric eclipsing system showing a slow apsidal motion with an important relativistic contribution as well as the rapid LITE caused by a third body orbiting with a very short period of 1.9 years. According to them, CG\,Aur probably belongs to an important group of other early-type and triple eclipsing systems with a very short third-body orbital period.

The apsidal motion in the eccentric eclipsing binaries has been used for decades to test the models of stellar structure and evolution. CG\,Aur analyzed here has some properties, which make the system an important "astrophysical laboratories" for studying the stellar structure and evolution. The lack of the multi-colour light curves makes CG Aur an interesting system for including it to our photometric programme. We followed the observations of the system and obtained the light curves in 2011 and 2012, and we discussed the light and colour variations. CG\,Aur is important due to not only being a member of the group of the triple systems, but also having an eccentric orbit.

\section{Observations}
The observations were acquired with a thermoelectrically cooled ALTA U+42 2048$\times$2048-pixel CCD camera attached to a 40-cm Schmidt-Cassegrain MEADE telescope at Ege University Observatory. Using exposure times of 100 s in B filter and 40 s in V, R and I filters, the BVRI-band observations were recorded over five nights in 2011 and four nights in 2012. Calibration images (bias frames and twilight sky flats) were taken intermittently during observations to correct for pixel to pixel variations on the chip. CCD observations were reduced as follows: Bias and dark frames were subtracted from the science frames and then corrected for the flat-fielding. These reduced CCD images were used to obtain the differential magnitudes of the program stars. We used GSC\,1857 833 and GSC\,1857 736 as a comparison and check stars shown in Figure 1, respectively. There was no variation observed in the brightnesses of the comparison star.

During the observations, we obtained one primary and one secondary times of minimum light. These minima times and their errors were determined using the method of \citet{Kwe56} and are presented in Table 1. In order to calculate the phases of the photometric data of CG\,Aur, the following linear ephemeris was used:

\begin{center}
\begin{equation}
JD~(Hel.)~=~24~55983.2658(6)~+~1^{d}.8048588(3)~\times~E.
\end{equation}
\end{center}

The V-light and the colour curves obtained in this study are shown in Figure 2. The shape of the light curves indicates that CG\,Aur is an Algol type binary and reveals that the primary minimum, which lasts $\sim$6 hours, is deeper than the secondary one. The mean depths of the eclipses in B, V, R, and I filters are 0$^{m}$.434, 0$^{m}$.390, 0$^{m}$.382, and 0$^{m}$.360 in the primary minimum and 0$^{m}$.168, 0$^{m}$.202, 0$^{m}$.218, and 0$^{m}$.235 in the secondary minimum,  respectively. The $B-V$, $V-R$ and $V-I$ colour curves of the system are also displayed in Figure 2. The system is slightly redder at the primary and bluer at the secondary minimum which is consistent with the spectral types of the components. 

\section{Light Curve Analysis}

Photometric analysis of CG\,Aur was carried out using the PHOEBE V.0.31a software \citep{Prs05}. The software uses the version 2003 of the Wilson-Devinney Code \citep{Wil71, Wil90}. The BVRI light curves were analysed simultaneously assuming the "detached" configuration. In the process of the computation, we initially adopted the following fixed parameters: the mean temperature of the primary component ($T_{1}$), the linear limb-darkening coefficients of $x_{1}$ and $x_{2}$ for various bands \citep{Van93}, the gravity-darkening exponents of $g_{1}$, $g_{2}$ \citep{Luc67} and the bolometric albedo coefficients of $A_{1}$, $A_{2}$ \citep{Ruc69}. The adjustable parameters commonly employed are the orbital inclination ($i$), the mean temperature of the secondary component ($T_{2}$), the potentials of the components ($\Omega_{1}$ and $\Omega_{2}$) and the monochromatic luminosity of the primary component ($L_{1}$). The third light ($L_{3}$) was used also as free parameter to check for the third light contribution. We used the $U-B$ and $B-V$ values given by \citet{Zak95} and determined the dereddened colours of the system as $(U-B)_{0}=0^{m}.06$, $(B-V)_{0}=0^{m}.24$. Then, we took JHK magnitudes of the system ($J=10^{m}.261$, $H=10^{m}.036$, $K=9^{m}.955$) from the 2MASS Catalogue \citep{Cut03}. Using these magnitudes, we derived dereddened colours as a $(J-H)_{0}=0^{m}.225$ and $(H-K)_{0}=0^{m}.081$ for the system. Using the calibrations given by \citet{Tok00}, we derived the temperature of the primary component as 7650 K and 7475 K depending on the UBV and JHK dereddened colours, respectively. Both of them indicate the same spectral range which is A5-F0. We adopted the mean temperature of the primary component as 7650 K for the light curve analyse. To find a photometric mass ratio, the solutions are obtained for a series of fixed values of the mass ratio from $q=0.4$ to 1.0 in increments of 0.1. The sum of the squared residuals ($\Sigma res^{2}$) for the corresponding mass ratios are plotted in Figure 3, where the lowest value of ($\Sigma res^{2}$) was found at about $q=0.7$. The photometric elements for the mass-ratio of 0.7 are listed in Table 2, and the corresponding light curves are plotted in Figure 4 as continuous lines. The Roche Lobe geometry of the system is displayed for the phase of 0.25 in Figure 5.

Although there is no available radial velocity curve of the system, we tried to estimate the absolute parameters of the components. Considering its spectral type, using the calibration of \citet{Tok00}, we estimate the mass of the primary component as approximately 1.78 $M_{\odot}$, and the mass of the secondary component is computed from the estimated mass ratio of the system. Using Kepler's third law, we calculate the semi-major axis ($a$), and also the mean absolute radii of the components. We can calculate the distance of the system by using primary and secondary component separately, via their photometric and absolute properties. We adopted bolometric corrections from \citet{Tok00}, while calculating the distance. Photometric properties of the components lead to an average distance of 460 pc. All the estimated absolute parameters are listed in Table 3. In Figure 6, we plot the components in the $log~(M/M_{\odot})$-$log~(R/R_{\odot})$ and $log~(T_{eff})$-$log~(L/L_{\odot})$ planes. The continuous and dotted lines represent the ZAMS and TAMS theoretical model developed by \citet{Gir00}. All the tracks are taken from \citet{Gir00} for the stars with $Z=0.02$. In the figure, the open circles represent the secondary component, while the filled circles represent the primary component. Both of the components locate in the main-sequence band, while the more massive component is more evolved.

\section{Summary}

We have obtained the multi-colour CCD photometry for the interesting eclipsing binary CG\,Aur. Based on these observations, we presented the first BVRI light curves of the system and analysed them to find the parameters obtained from the orbital solution, considering also the Roche configuration of the system. The results allow us to draw the following conclusions.

$\bullet$ The physical and geometrical parameters of the components have been derived. Orbital parameters indicate that CG\,Aur is a detached binary system with a little temperature difference of approximately 1000 K.

$\bullet$ Our photometric model describes CG\,Aur as an Algol type eclipsing binary in which the more massive and hotter primary component is the larger one.

$\bullet$ In order to discuss the present evolutionary status of the components of CG\,Aur, both of them were plotted on the $log~(M/M_{\odot})$-$log~(R/R_{\odot})$ and $log~(T_{eff})$-$log~(L/L_{\odot})$ planes. The components of the system located inside the Main-Sequence band. The primary component is slightly evolved from the Zero-Age Main-Sequence (ZAMS), while the secondary component is still very close to ZAMS. CG\,Aur is a detached eclipsing binary and its components are Main-Sequence dwarfs, the evolutionary states of the components of the system could be estimated from single-star evolutionary models. We compared the physical parameters of the system with those inferred from the evolutionary tracks for single stars of masses in the range of 1.8 and 1.3 $M_{\odot}$, taken from \citet{Gir00} for solar composition. There is a good agreement between the estimated masses and evolutionary masses for the components of the system.

$\bullet$ The detailed $O-C$ study of the system were presented by \citet{Wol11}. In this study, we obtained a small amount of the third light contribution to the total light from the light curve solution of the system for the first time. According to our solution, the luminosity fraction of the third body was found maximum in I filter, while the least contribution was determined in B filter, which indicates its cool nature. The spectral type of the third body should be late type.

$\bullet$ As said by \citet{Wol11}, CG\,Aur is an interesting triple and eccentric eclipsing system showing the slow apsidal motion with the important relativistic contribution as well as the rapid LITE caused by a third body orbiting with the very short period of 1.9 years. CG\,Aur is shown to be an Algol-type system of special interest because of its nature. The system belongs to the important group of other early-type and triple eclipsing systems with a very short third-body orbital period (e.g. IM\,Aur, IU\,Aur, FZ\,CMa, AO\,Mon) as said by \citet{Wol11}, and also it has an eccentric orbit. CG\,Aur is important due to not only being a member of the group of the triple systems, but also having an eccentric orbit, while short period binaries with periods less than a week generally have circular orbits. It is recommended that more attention should be paid to this system.

$\bullet$ In future work, spectroscopic observations should be made to obtain radial velocity curves, which will allow a better discussion of the absolute dimensions of the components and the evolutionary status of CG\,Aur. New timings of this eclipsing binary are also necessary to improve the LITE parameters of the system.

\section*{Acknowledgment} The author acknowledge generous allotments of observing time at the Ege University Observatory. We also thank the referee for useful comments that have contributed to the improvement of the paper.

\clearpage

\begin{figure}
\hspace{2.2 cm}
\includegraphics[width=9cm]{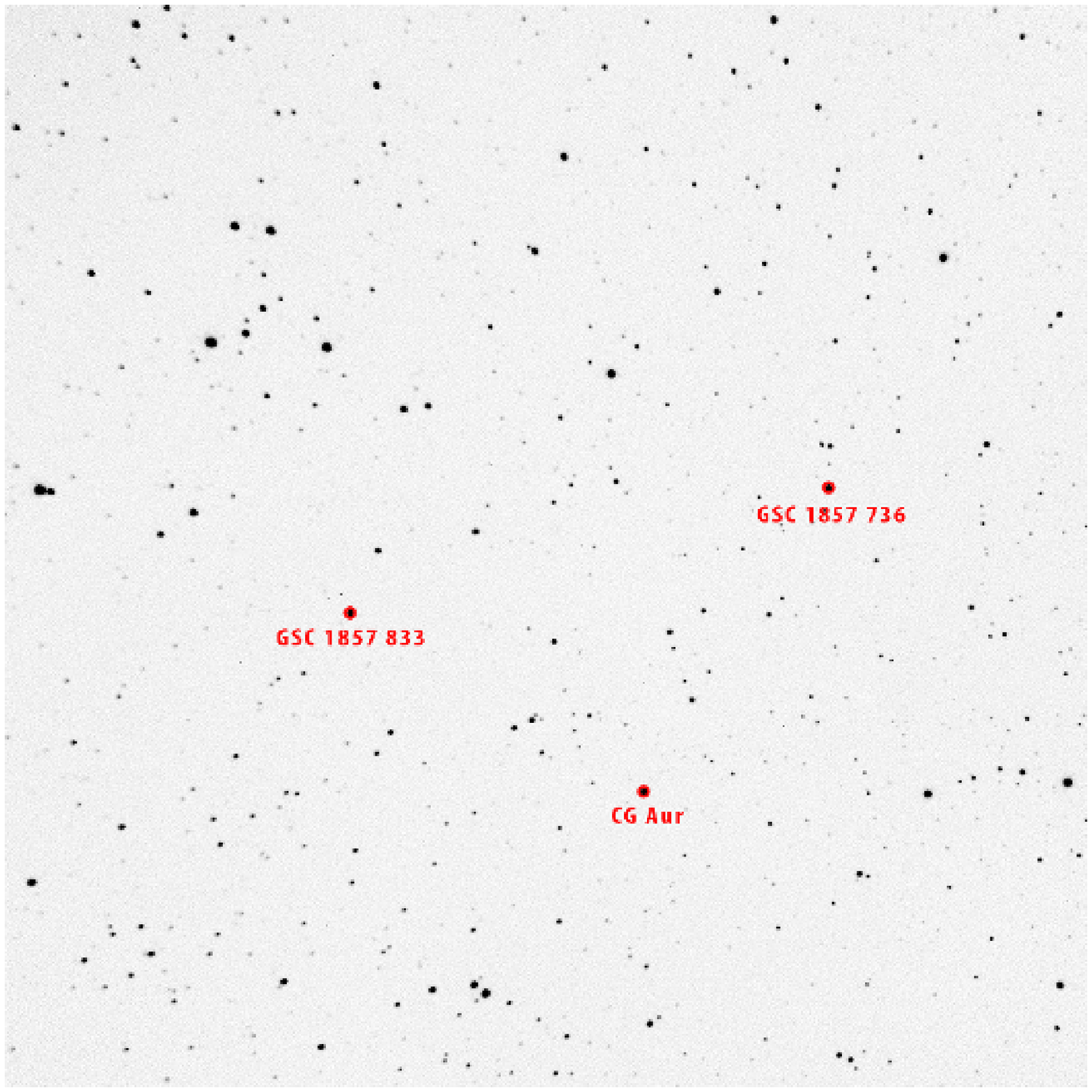}
\caption{The program star CG\,Aur and comparison and check stars on the sky plane, which is in size of $23^{\prime}.27\times23^{\prime}.27$.}
\label{Fig.1}
\end{figure}

\begin{figure}
\hspace{3.5 cm}
\includegraphics[width=20cm]{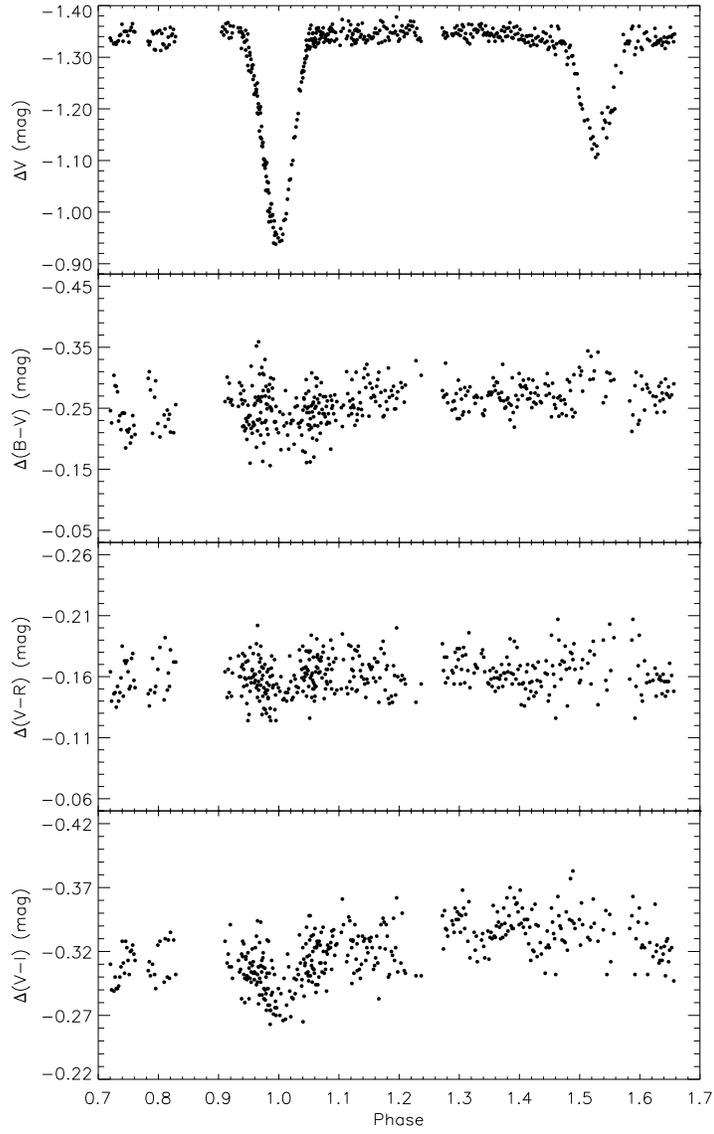}
\vspace{0.01cm}
\caption{CG\,Aur's V-light and colour curves.}
\label{Fig.2}
\end{figure}

\begin{figure}
\hspace{3.0 cm}
\includegraphics[width=15cm]{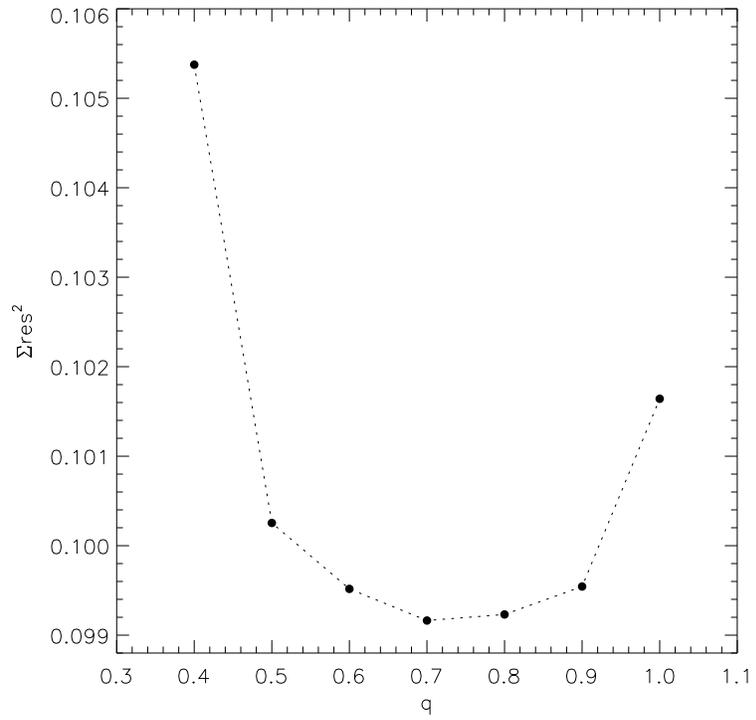}
\vspace{0.01cm}
\caption{The variation of the sum of weighted squared residuals versus mass ratio.}
\label{Fig.3}
\end{figure}

\begin{figure}
\hspace{3.5 cm}
\includegraphics[width=20cm]{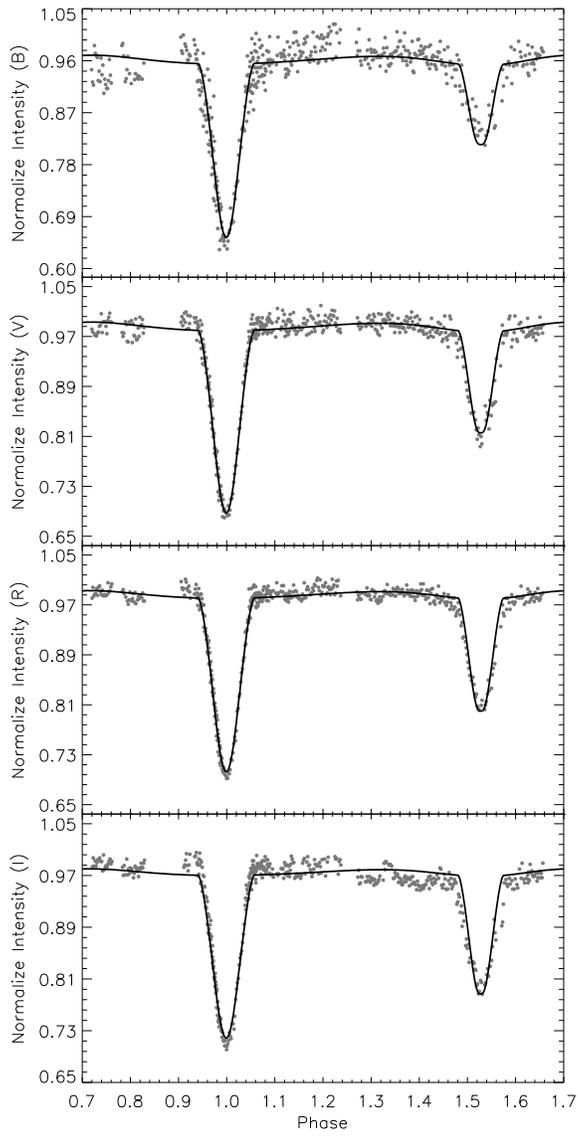}
\vspace{0.01cm}
\caption{CG\,Aur's light curves observed in BVRI bands and the synthetic curves derived from the light curve solutions in each band.}
\label{Fig.4}
\end{figure}

\begin{figure}
\includegraphics[width=15cm]{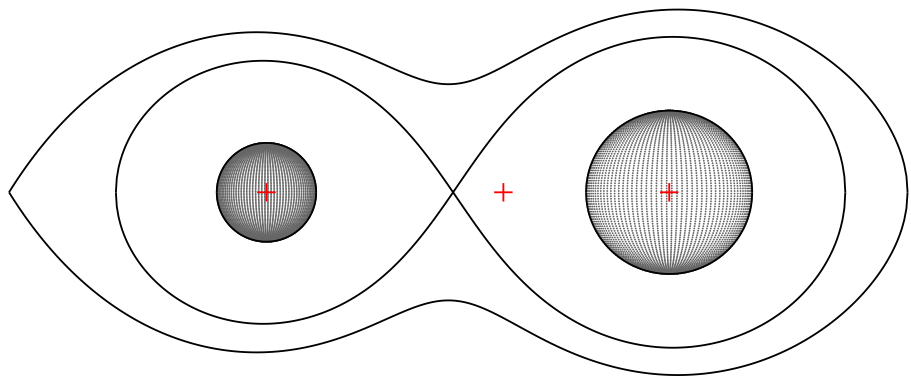}
\vspace{-8.00cm}
\caption{CG\,Aur's Roche geometry, while the system is at the phase of 0.25.}
\label{Fig.5}
\end{figure}

\begin{figure}
\hspace{3.2 cm}
\includegraphics[width=17cm]{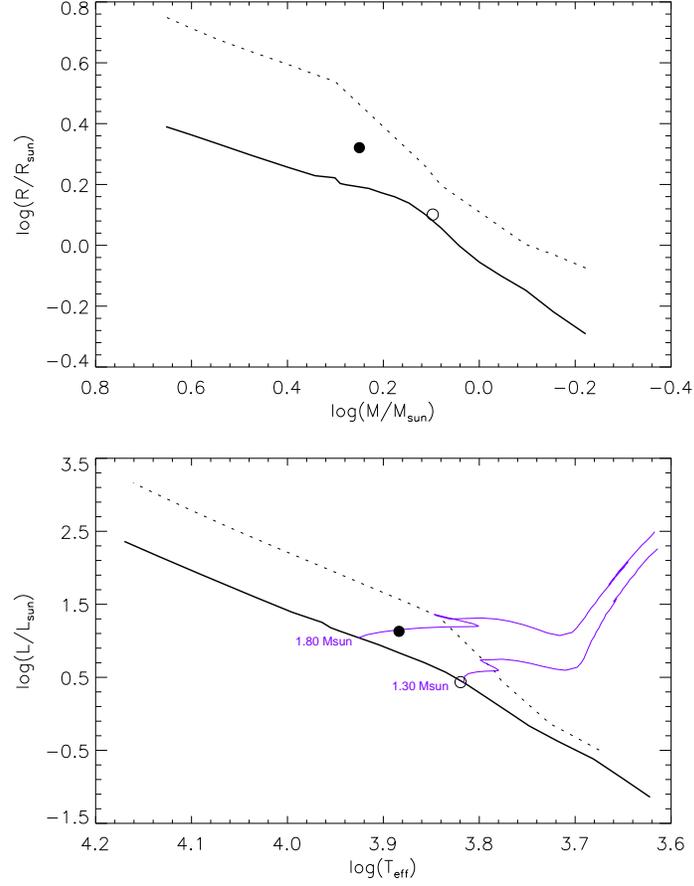}
\vspace{0.01cm}
\caption{The places of the components of CG\,Aur in the $log~(M/M_{\odot})$-$log~(R/R_{\odot})$ plane (upper panel) and $log~(T_{eff})$-$log~(L/L_{\odot})$ plane (bottom panel). In the figures, the filled circles represent the primary, while open circles represent the secondary component. The continuous and dotted lines represent the ZAMS and TAMS theoretical model developed by \citet{Gir00}. All the tracks taken from \citet{Gir00} are derived for the stars with $Z=0.02$.}
\label{Fig.6}
\end{figure}

\clearpage

\begin{table*}
\centering
\caption{The times of minimum light for CG\,Aur.}
\vspace{0.3cm}
\begin{tabular}{cccc}
\hline\hline
HJD (24 00000 +)	&	Sigma	&	Type	&	Filter	\\
\hline							
55983.2661	&	0.0010	&	I	&	B	\\
55983.2655	&	0.0006	&	I	&	V	\\
55983.2659	&	0.0004	&	I	&	R	\\
55983.2656	&	0.0004	&	I	&	I	\\
56002.2782	&	0.0010	&	II	&	B	\\
56002.2668	&	0.0018	&	II	&	V	\\
56002.2732	&	0.0020	&	II	&	R	\\
56002.2682	&	0.0008	&	II	&	I	\\
\hline
\end{tabular} 
\end{table*}

\begin{table*}
\centering
\caption{The parameters obtained from the light curve analysis.}
\vspace{0.3cm}
\begin{tabular}{lrlr}
\hline\hline
Parameter	&	Value	& Parameter	&	Value\\
\hline			
$T_{0}$ 	&	 24 50014.6248	&	$P$ (day) 	&	 1.8048588 \\
$q$ 	&	0.7	&	$i$ ($^\circ$) 	&	 87.74$\pm$0.01 \\
$e^{1}$ 	&	0.124	&	$w$ ($^\circ$) 	&	 291.05$\pm$0.01 \\
$T_{1}$ (K) 	&	7650	&	$T_{2}$ (K) 	&	 6600$\pm$27 \\
$\Omega_{1}$ 	&	 5.617$\pm$0.001	&	$\Omega_{2}$ 	&	 6.850$\pm$0.002 \\
L$_{1}$/L$_{T}$ $(B)$ 	&	 0.862$\pm$0.015	&	L$_{3}$/L$_{T}$ $(B)$ 	&	 0.0007$\pm$0.0002 \\
L$_{1}$/L$_{T}$ $(V)$ 	&	 0.836$\pm$0.014	&	L$_{3}$/L$_{T}$ $(V)$ 	&	 0.0020$\pm$0.0002 \\
L$_{1}$/L$_{T}$ $(R)$ 	&	 0.814$\pm$0.014	&	L$_{3}$/L$_{T}$ $(R)$ 	&	 0.0031$\pm$0.0002 \\
L$_{1}$/L$_{T}$ $(I)$ 	&	 0.792$\pm$0.012	&	L$_{3}$/L$_{T}$ $(I)$ 	&	 0.0080$\pm$0.0002 \\
$g_{1}$, $g_{2}$ 	&	 0.32, 0.32	&	$A_{1}$, $A_{2}$ 	&	 0.5, 0.5 \\
Phase shift 	&	 0.0146$\pm$0.0002	&	$x_{1,bol}$, $x_{2,bol}$ 	&	 0.522, 0.481 \\
$x_{1,B}$, $x_{2,B}$ 	&	 0.596, 0.656	&	$x_{1,R}$, $x_{2,R}$ 	&	 0.421, 0.434 \\
$x_{1,V}$, $x_{2,V}$ 	&	 0.518, 0.532	&	$x_{1,I}$, $x_{2,I}$ 	&	 0.329, 0.351 \\
$<r_{1}>$ 	&	 0.209$\pm$0.001	&	$<r_{2}>$ 	&	 0.126$\pm$0.001 \\
\hline
\end{tabular}
\begin{list}{}{}															
\item[$^{1}$]{\small Taken from \citet{Wol11}}																											
\end{list} 
\end{table*}

\begin{table*}
\centering
\caption{The estimated absolute parameters derived for CG\,Aur.}
\vspace{0.3cm}
\begin{tabular}{lccc}
\hline\hline
Parameter	&	Primary	&&	Secondary	\\
\hline					
Mass ($M_{\odot}$)	&	1.78	&&	1.25	\\
Radius ($R_{\odot}$)	&	2.09	&&	1.26	\\
Luminosity ($L_{\odot}$)	&	13.49	&&	2.72	\\
$M_{bol}$ (mag)	&	1.92	&&	3.65	\\
$log~(g)$	&	4.05	&&	4.33	\\
$d$ (pc)	&	& 460	&		\\
\hline
\end{tabular} 
\end{table*}

\end{document}